\documentclass[12pt]{iopart}
\usepackage{epsfig}
\usepackage{iopams}
\newcommand{\half}{\frac{1}{2}}
\newcommand{\sqnbar} {\sqrt{\bar{n}}}
\newcommand{\nbar} {\bar{n}}

\newcommand{\ket}[1]{ {| #1 \rangle} }
\newcommand{\bra}[1]{ {\langle #1 |} }
\newcommand{\abs}[1]{ {| #1 |}}
\newcommand{\braket}[2]{ {\langle #1  |#2\rangle} }

\begin{document}

\title{Entanglement of superconducting charge qubits by
homodyne measurement}
\author{D A Rodrigues$^1$, C E A Jarvis$^2$, B
L Gy\"{o}rffy$^2$, T P Spiller$^3$ and J F Annett$^2$}%
\address{$^1$ School of Physics and Astronomy, University of
Nottingham, Nottingham NG7 2RD, United Kingdom}%
\address{$^2$ H H Wills Physics Laboratory, University of Bristol, Bristol BS8 1TL, United
Kingdom}%
\address{$^3$ Hewlett Packard Laboratories, Filton Road, Bristol, BS34 8QZ, United Kingdom}%
\ead{$^1$ denzil.rodrigues@nottingham.ac.uk} %
\ead{$^2$ catherine.jarvis@bristol.ac.uk}

% ----------------------------------------------------------------

\begin{abstract}

We present a scheme by which projective homodyne measurement of a microwave
resonator can be used to generate entanglement between two
superconducting charge qubits coupled to this resonator. The
non-interacting qubits are initialised in  a product of their ground
states, the resonator is initialised in a coherent field state, and
the state of the system is allowed to evolve under a rotating wave
Hamiltonian. Making a homodyne measurement on the resonator at a
given time projects the qubits into an state of the form
$(\ket{gg}+e^{-i\phi}\ket{ee})/\sqrt{2}$. This protocol can produce
states with a fidelity as high as required, with a probability
approaching 0.5. Although the system described is one that can be
used to display revival in the qubit oscillations, we show that the
entanglement procedure works at much shorter timescales.

\end{abstract}
%\submitto{NJP}

\maketitle
% ----------------------------------------------------------------

\section{Introduction}

Central to the construction of any useful Quantum Information device
will be an array of quantum systems  whose collective quantum state
can be prepared, measured or otherwise manipulated \cite{Vincenzo}.
For the case of two-level systems---qubits---and in the solid state
arena, small superconducting grains coupled to bulk superconductors
via Josephson junctions, often referred to as Cooper-pair boxes,
have been shown to be promising candidates for playing the role of
the qubits\cite{Nakamura}. Quantum state control has been
demonstrated in single qubit devices through
coherent or Rabi oscillations,
observed with readout schemes consisting of either a single electron
transistor (SET)\cite{Nakamura} or quantronium\cite{Vion} circuit
coupled to the qubit being measured.  Corresponding single qubit
quantum state control results have also been obtained in
superconducting flux qubit devices\cite{Chiorescu}.

More recently, the coupling of such superconducting qubit devices
has also been achieved\cite{Pashkin03,Berkley03,McDermott05}.  Such
coupling generates entangled two-qubit states, which is both of
vital importance to possible quantum computing
applications\cite{Vincenzo} , and also of interest in testing the
fundamental limits of quantum mechanics in macroscopic
objects\cite{Leggett02,leggett07}. The two charge qubits of Pashkin
\emph{et al.}\cite{Pashkin03} were capacitively coupled, and each
qubit independently measured with SET readouts. The resulting
two-body density matrix, was consistent with the existence of
non-classical correlations between the two qubits, although
decoherence limited the fidelity of the CNOT gate which was
attempted in the two qubit device\cite{Yamamoto}. Entanglement
between superconducting  flux qubit devices was reported by Berkley
\emph{et al.}\cite{Berkley03}. A more recent
experiment\cite{McDermott05} showed antiphase oscillation of a
two-qubit system, and detailed quantum state tomography established
entanglement with 87\% fidelity between two macroscopic
superconducting charge-phase qubit devices\cite{Steffen}.
Significant progress has also been made coupling superconducting
charge\cite{Wallraff,Wallraff2,Schuster} and flux\cite{Chiorescu2}
qubits to a quantum resonator mode, demonstrating effects such as
the AC-Stark effect and resolving individual photon number states in
the resonator. As well as being an essential part of any quantum computing protocol,  the generation of entanglement is necessary for the demonstration of a Bells' inequality violation\cite{Bell}, which would prove unequivocally that the system cannot be described classically\cite{Marchese}.

In all these experiments the coupling of the superconducting qubit
to its measurement device has been shown to be one of the central
aspects controlling its decoherence.  Very long relaxation times
were observed in the experiments of Wallraff {\it et
al.}\cite{Wallraff,Wallraff2,Schuster} in which the Cooper-pair box
was coupled to a microwave stripline resonator, but not to any other
measurement device. Measurement of a resonator dispersively coupled
to a superconducting qubit has thus been shown to provide an
excellent `readout' system for the superconducting
qubit\cite{Wallraff2,Lupascu}. Quantum resonator modes therefore
have real potential for both coupling and readout of superconducting
qubits.

In this paper we present a measurement protocol by which
entanglement may be generated between two such superconducting
qubits, which are coupled only though a common stripline resonator.
We show that projective homodyne measurement on the resonator field may perform
a projective measurement on the coupled qubit/field system,
resulting in the generation of entangled states of the two qubits.
It is interesting that there is assumed to be no direct coupling
between the qubits themselves, and it is the act of projective
measurement on the field which creates the entangled state. Our
scheme for generation of entanglement is therefore analogous to the
Knill, Laflame, Milburn or `KLM' protocol\cite{KLM} to generate
entanglement of photons in linear quantum optics through projective
measurement. In the same way, our scheme uses measurement to
generate entanglement between qubits which have no other direct
qubit-qubit interaction. The analysis presented here will be relevant for other physical systems consisting of two-level systems coupled via an harmonic mode such as atom-chips or nanomechanical systems.

Our work builds upon the  extensive theoretical study of Meunier
\emph{et al.}\cite{Meunier} of collapse and revival phenomena of two
level quantum systems coupled to a single quantized radiation mode.
As is stressed by Meunier \emph{et al.}\cite{Meunier} the rich
variety of quantum phenomena displayed by the  `one qubit one mode'
system is well known from Quantum Optics where the role of the two
level system is played by a two level (Rydberg)
atom\cite{ShoreKnight93}. A particularly striking example of these
is the collapse and subsequent revival of the initial Rabi
oscillations in the atom (qubit) section of the Hilbert space when
the initial radiation field is in a coherent state $\ket{\alpha}$.
Remarkably, as was first noticed by Gea-Banacloche \cite{Ban91}, at
times $t$, between the collapse time $t_c$ and revival time $t_r$,
when there are no Rabi oscillations, it is the radiation field part
of the wave-function which manifests complex quantum behaviour.
Following these insights we have studied the time evolution of the
two qubit system coupled to a radiation mode. We show that simple
projective homodyne measurements on the radiation field, at times between
collapse and revival of multi qubit Rabi oscillations, may be used
to obtain interesting entangled states of the two qubits.

The paper is structured as follows. In section 2 we introduce the
two qubit version of the Jaynes-Cummings
model\cite{JaynesCum,TavisCum} and, in the interest of clarity,
recall the current state of understanding of the complex evolution
with time of a combined qubit and field state. We also discuss the
expected results of a homodyne measurement on the field variable of
our system. The main results of this paper are developed in section
3 where we describe and analyse a simple experimental protocol which
involves homodyne measurements on the field variable and results in
heralded maximally entangled states of the qubits. Our general
conclusions are presented in section 4.

\section{Jaynes-Cummings model for two qubits coupled to a resonator}

%         Figure of setup?

The salient features of one qubit coupled to a single mode of the
radiation field can be described approximately by the
Jaynes-Cummings model\cite{JaynesCum}, much studied in quantum
optics. This model has generated a great deal of interest in the
past as it exhibits interesting behaviour\cite{ShoreKnight93} and it
has been used to describe quantum correlation.

In this paper we consider two charge qubits each coupled
capacitively to a stripline resonator using the two qubit
Jaynes-Cummings model\cite{TavisCum}. If the qubits are placed at an
antinode of the fundamental harmonic mode of the resonator, then we
can describe the system as a pair of two-level systems coupled to a
simple harmonic oscillator. The charging energy of the qubits and
their coupling to the resonator can be controlled by the application
of magnetic and electric fields\cite{Nakamura}. If these are tuned
so that the qubits are close to resonance with the field we can
describe the system using a rotating wave Hamiltonian,

\begin{eqnarray}\label{eq:ham}
\hat{H}&=&\hbar\omega\left(\hat{a}^{\dag}\hat{a}+\frac{1}{2}\right)+\left(E_{1}\hat{\sigma}_{1}^{z}+E_{2}\hat{\sigma}_{2}^{z}\right)+\hbar
\sum_{i=1}^{2}
\lambda_{i}\left(\hat{a}\hat{\sigma}_{i}^{+}+\hat{a}^{\dag}\hat{\sigma}_{i}^{-}\right)
\end{eqnarray}
where
\begin{eqnarray}
\hat{\sigma}^{z}=\ket{e}\bra{e}-\ket{g}\bra{g},\
\hat{\sigma}^{+}=\ket{e}\bra{g},\ \hat{\sigma}^{-}=\ket{g}\bra{e},
\end{eqnarray}
$\hat{a}^{\dag}$ and $\hat{a}$ are the creation and annihilation
operators of photons with frequency $\omega$, $E_{1,2}$ are the
charging energies of the qubits, and $\lambda_{1,2}$ the
resonator-qubit coupling terms. We consider the system to be exactly
on resonance, for simplicity, and also assume that the Cooper pair
boxes have equal charging energy, $(E_{1,2}=E)$ and coupling to the
resonator, $(\lambda_{1,2}=\lambda)$. In the rotating wave
approximation, we notice that states with $m$ excited qubits and $n$
photons only couple to states with $m'+n'=m+n$. This means that we
can describe the system as an infinite set of non-interacting four
level sub-systems all labeled by $n+m$. We can find the solution of
each of these sub-systems analytically, and then sum over these to
obtain the state of the whole system.

For an initial state,
\begin{equation}
\ket{\psi(t)}=\ket{gg}\sum_{n=0}^{\infty}C_{n}\ket{n}
\end{equation}
we obtain,
\begin{eqnarray}\label{eq:ansoln}
\ket{\psi(t)}&=& \sum_{n=0}^\infty\exp\left(-i \frac{\omega}{2}
(2n-1)t\right)C_{n}\nonumber\\ & &
\Bigg[\frac{\sqrt{n(n-1)}}{2n-1}\left(\cos\left(\lambda
t\sqrt{2(2n-1)}\right)-1\right)\ket{ee, n-2} \nonumber\\ &-& i
\frac{\sqrt{n}}{\sqrt{2(2n-1)}}\sin\left(\lambda t\sqrt{2(2n-1)}\right)(\ket{eg, n-1}+\ket{ge, n-1})\nonumber\\
&+& \frac{1}{2n-1}\left(n\cos\left(\lambda
t\sqrt{2(2n-1)}\right)+(n-1)\right)\ket{gg,
n}\Bigg]\nonumber\\
\end{eqnarray}
where the state $\ket{eg, n}$ is the tensor product of a Fock state
$\ket{n}$ of the photons and the state of the qubits which is either
excited ($e$) or ground ($g$) for each qubit. For brevity we only
discuss the solution for the case where the qubits both start in
their ground state, but we note that the solution with a general
initial condition is generically similar. The initial state of the
field is determined by the coefficients $C_n$, which for a coherent
state $\ket{\alpha}$ are given by,
\begin{equation}\label{eqn:coherent}
C_n =e^{-\abs{\alpha}^{2}/2}\frac{\alpha^n}{\sqrt{n!}},
\end{equation}
with the phase, $\theta$, and average occupation, $\bar{n}$, of the
coherent state determined by $\alpha=\sqrt{\bar{n}}e^{-i\theta}$.

\begin{figure}\center{
\epsfig{file=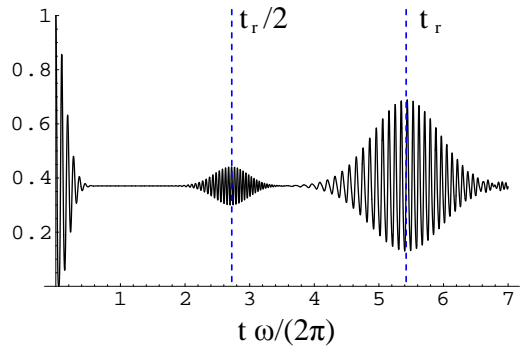, height=5cm}} \caption{Revival of the
initial state of the qubit system. The curve shows the probability
that the qubits are in the state $\ket{gg}$ after tracing out the
resonator. The initial qubit oscillations rapidly decay, indicating
the qubits are in a mixed state. The revival of the oscillations
indicates that the information about the qubit states has been
transferred back to the qubits from the resonator. The main revival
occurs at $t_r=2\pi\sqrt{\bar{n}}/\lambda$, with a partial revival
at $t_r/2$. We have chosen $\omega=\lambda=1$ and $\bar{n}=30$.
\label{fig:revival}}
\end{figure}

In spite of its relative simplicity equation (\ref{eq:ansoln})
describes a wide range of interesting phenomena. A much studied
example of these is the collapse and revival of Rabi oscillations.
In this paper we shall be concerned with phenomena at times where
there are no Rabi oscillations. To illustrate the generic behaviour
of $\ket{\psi(t)}$ in equation (\ref{eq:ansoln}) we shall now
comment on the qubit and the photon sector separately. If we
consider the state of the qubits after tracing out the field state,
the probability of the qubits being in the state $\ket{gg}$ is,
\begin{equation}
P_{gg}(t)=\bra{gg}\rho^{Q}(t)\ket{gg}=\sum_{n=0}^{\infty}\abs{\braket{gg,n}{\psi(t)}}^2
\end{equation}
where $\rho^Q(t)$ is the reduced density matrix at a certain time
for the qubits when the field has been traced out. In agreement with
Iqbal \emph{et al.}\cite{Iqbal}, we find the results depicted in
figure \ref{fig:revival}. Evidently we see that initial oscillations
of the qubit states decay rapidly, after which the qubit is in a
mixed state. After a period of time, the oscillations revive,
indicating the information about the qubits' initial state has
returned to the qubits from the field. We see a larger revival at
the revival time, $t_r=2\pi \sqnbar /\lambda$, which is proceeded by
a smaller revival at $t_r/2$. The smaller revival is not seen for
the one qubit case and this is due to the extra frequency introduced
by the addition of the second qubit.

\begin{figure}[h!]\center{
\epsfig{file=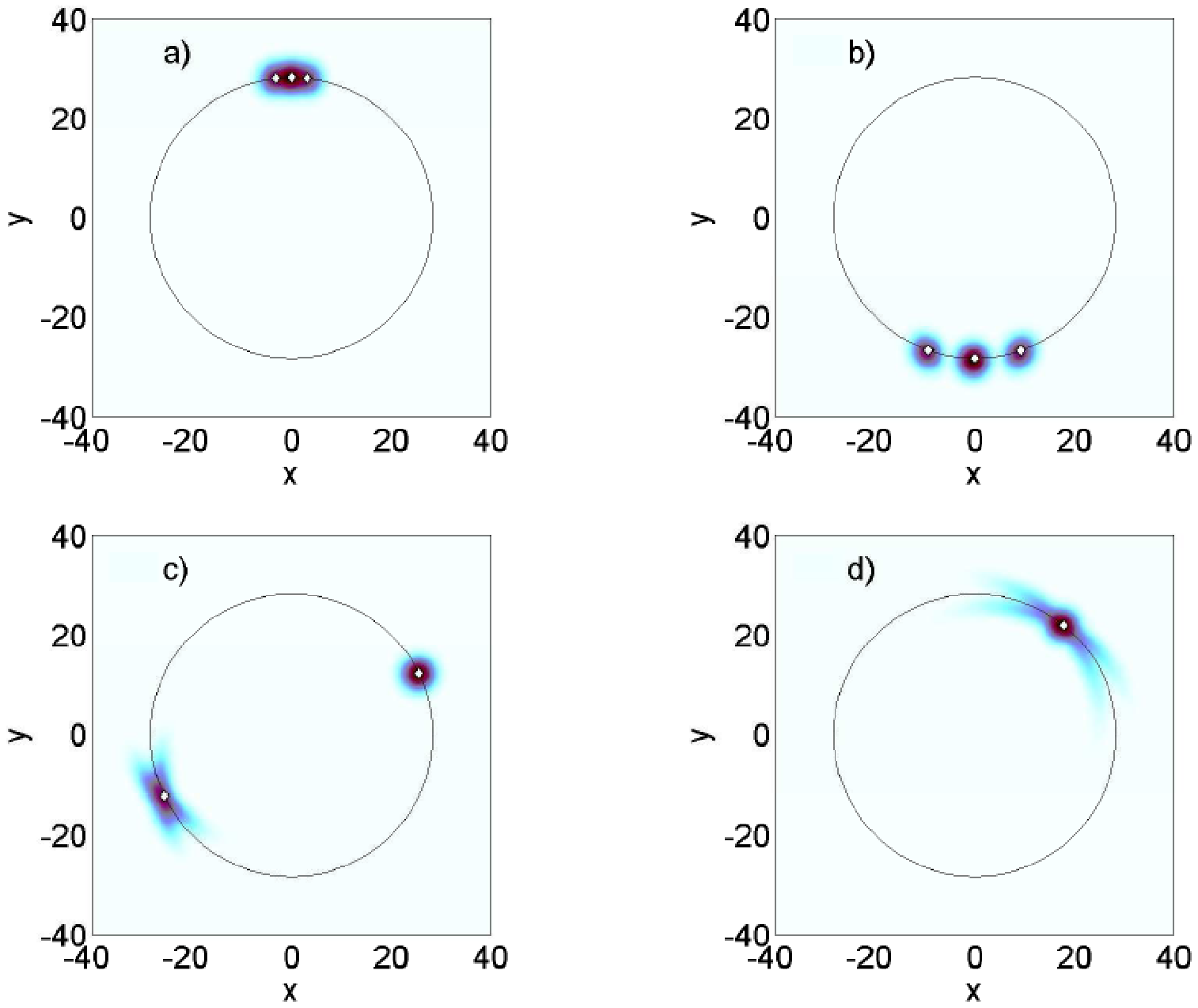}}
\caption{Phase-space plots of the Q-function for four different
times, where the dimensionless $x=\alpha+\alpha^*$ represents the
electric field and $y=(\alpha-\alpha^*)/i$ represents the magnetic
field. Also plotted are white diamonds corresponding to the
evolution of the field states in equation (\ref{eq:Phik}). (a)
$t=\pi/(2\omega)$: The phase space close to $t=0$, when the peaks
are not well separated, (b) $t=3\pi/(2\omega)$: After a period of
time the sub-distributions, evolving with different frequencies,
separate and can easily be distinguished, (c) $t=t_{r}/2$: The time
at which there is the first occurrence of spontaneous revival, (d)
$t=t_{r}$: Location of the states at the second main revival peak.
For all the diagrams $\lambda=\omega=1$ and
$\bar{n}=200$.\label{fig:qblobs}}
\end{figure}

So far we have studied the time evolution of the reduced density
matrix of the qubits. To investigate the radiation field it is
useful to calculate the Q function \cite{GardinerZoller},
\begin{equation}\label{eq:qfunc}
Q(\alpha,t)=\bra{\alpha}\rho^{F}(t)\ket{\alpha}
\end{equation}
where $\rho^{F}(t)$ is the reduced density matrix at a certain time
for the radiation field when the qubits have been traced out. The
values of $Q(\alpha,t)$ in the complex alpha plane at fixed times
are shown in figure \ref{fig:qblobs}.

The Q function shows a set of three `blobs', each representing a
mesoscopic wavepacket of the cavity field. As a function of time the
three wavepackets move around the complex plane and follow the
circular path of radius $\sqrt{\bar{n}}$. Although the wavepackets
all begin in the same place, they evolve with different frequencies,
so the states begin to separate. After a period of time, depending
on the differences between the frequencies, the different
sub-distributions are separated by more than their diameter and can
be easily distinguished, (figure \ref{fig:qblobs}b).

To elucidate the connection between the collapse and revival in the
qubit part of the Hilbert space and the peaks in the $Q(\alpha,t)$
distribution depicted in figure \ref{fig:qblobs} we recall briefly
the results of Meunier \emph{et al.}\cite{Meunier} obtained in the
large $\nbar$ limit. Generalising the one qubit result of
Gea-Banacloche for the two qubit case they found that
$\ket{\psi(t)}$ can be written as a superposition of Gea-Banacloche
states described below.

\begin{eqnarray}\label{eq:productstates}
\ket{\psi(t)}&=&\sum_{k=-1}^{1}
\ket{D_{k}(t)}\otimes\ket{\Phi_{k}(t)}
\end{eqnarray}
where k=-1, 0, or 1, $\ket{D_{k}(t)}$ is a state of the qubits and
$\ket{\Phi_{k}(t)}$ is the state of the field.
\begin{eqnarray}\label{eq:dminusone}
\ket{D_{-1}(t)}&=&\frac{1}{4}(e^{-2i(\omega-\frac{\lambda}{\sqrt{\bar{n}}})t}
e^{-2i\theta}\ket{ee}-e^{-i(\omega-\frac{\lambda}{\sqrt{\bar{n}}})t}e^{-i\theta}(\ket{eg}+\ket{ge})+\ket{gg})\nonumber\\
& &\\
\label{eq:dzero}
\ket{D_{\ 0}(t)}&=& \half( \ket{gg}+e^{-2i(\theta+\pi/2+\omega t)}\ket{ee})\\
\label{eq:dplusone}
 \ket{D_{\ 1}(t)}&=&\frac{1}{4}(e^{-2i(\omega+\frac{\lambda}{\sqrt{\bar{n}}})t}
e^{-2i\theta}\ket{ee}+e^{-i(\omega+\frac{\lambda}{\sqrt{\bar{n}}})t}e^{-i\theta}(\ket{eg}+\ket{ge})+\ket{gg})\nonumber\\
& &\\
\label{eq:Phik}
\ket{\Phi_{k}(t)}&=&e^{-ik\lambda\sqrt{\bar{n}}t}\ket{e^{-i(\omega+k\frac{\lambda}{\sqrt{\bar{n}}})t}\alpha}
\end{eqnarray}
The state of the field in each Gea-Banacloche state
$\ket{D_{k}(t)}\otimes\ket{\Phi_{k}(t)}$ is one of three coherent
field states $|\Phi_k(t)\rangle$, each of which has a phase that
evolves with a frequency given by
$\omega+k\frac{\lambda}{\sqrt{\bar{n}}}$. Each field state
corresponds to a particular qubit state $\ket{D_{k}(t)}$, as given
by equations (\ref{eq:dminusone})-(\ref{eq:dplusone}). In
particular, we note that qubit state $\ket{D_{\ 0}}$ corresponding
to the field state $|\Phi_0(t)\rangle$ is a superposition of
$\ket{gg}$ and $\ket{ee}$ only, and has no components of the states
$\ket{ge}$ or $\ket{eg}$, a fact that we shall exploit to produce
our entanglement protocol.

%In this large $\nbar$ limit we can consider the system as a set of
%harmonic oscillators, each qubit state corresponding to a set of two
%or three oscillators with different frequencies. Consider the term
%in equation \ref{eq:ansoln} representing the state $\ket{gg}$. In
%the limit $n\gg1$, this term has components with three different
%frequencies, $\omega$ and $\omega\pm 2\lambda /\sqnbar$. These
%correspond to the three frequencies of the radiation field shown in
%equation \ref{eq:Phik} i.e. $\omega +k\lambda/\sqrt{\bar{n}}$ where
%k=0, $\pm$ 1. The term representing $\ket{ee}$ also contains these
%frequencies, but the terms representing $\ket{eg}+\ket{ge}$ only
%contain the frequencies $\omega\pm 2\lambda /\sqnbar$. This explains
%the composition of the qubit states represented in equations
%\ref{eq:dminusone}-\ref{eq:dplusone}.

The Gea-Banacloche states can be clearly seen in figure
\ref{fig:qblobs}, where the three `blobs' represent the states
$|\Phi_k(t)\rangle$, and we see that the frequency at which these
sub-distributions move around the complex plane is determined by the
frequencies in equation (\ref{eq:Phik}). The well known collapse and
revival phenomena, seen in figure \ref{fig:revival}, can now be
understood in terms of the evolution of the Gea-Banacloche states.
The Rabi oscillations collapse when the field states are well
separated (figure \ref{fig:qblobs}(b)). At the large revival peak
(figure \ref{fig:qblobs}(d)) all the distributions overlap in the
complex plane, which occurs when all the field states are in phase.
The time this occurs is determined by the condition $\exp(i\omega
t)=\exp(i(\omega +\lambda/\sqrt{\bar{n}})t)=\exp(i(\omega
-\lambda/\sqrt{\bar{n}})t)$. This condition is fulfilled by
$t_{r}=2\pi \sqrt{\bar{n}}/\lambda$. At an earlier time we observe a
smaller revival peak, when two of the states are in phase (figure
\ref{fig:qblobs}(c)). This occurs when $\exp(i(\omega
+\lambda/\sqrt{\bar{n}})t)=\exp(i(\omega
-\lambda/\sqrt{\bar{n}})t)$, namely at $t=t_{r}/2$.

\begin{figure}[h!]\center{
\epsfig{file=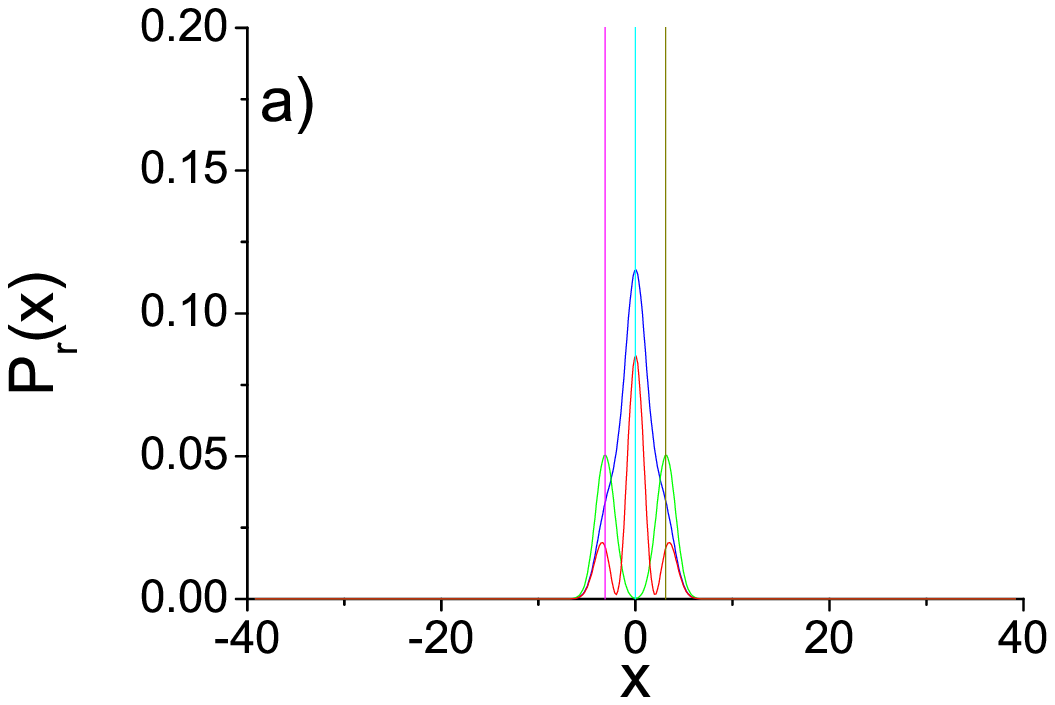, height=6cm}\epsfig{file=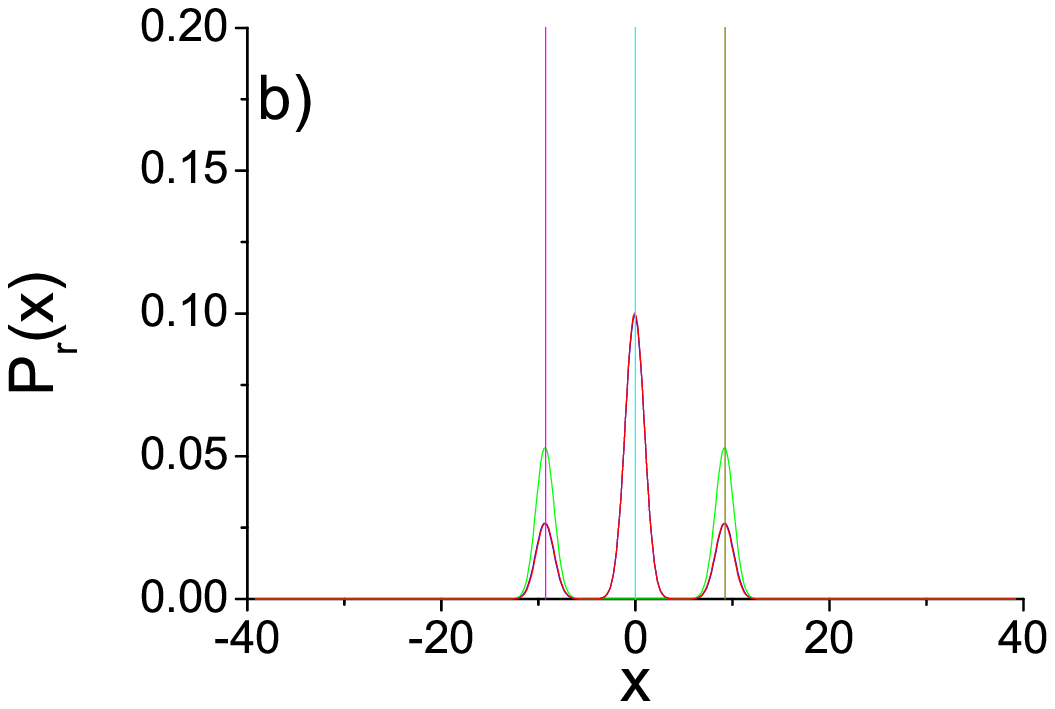,
height=6cm} \epsfig{file=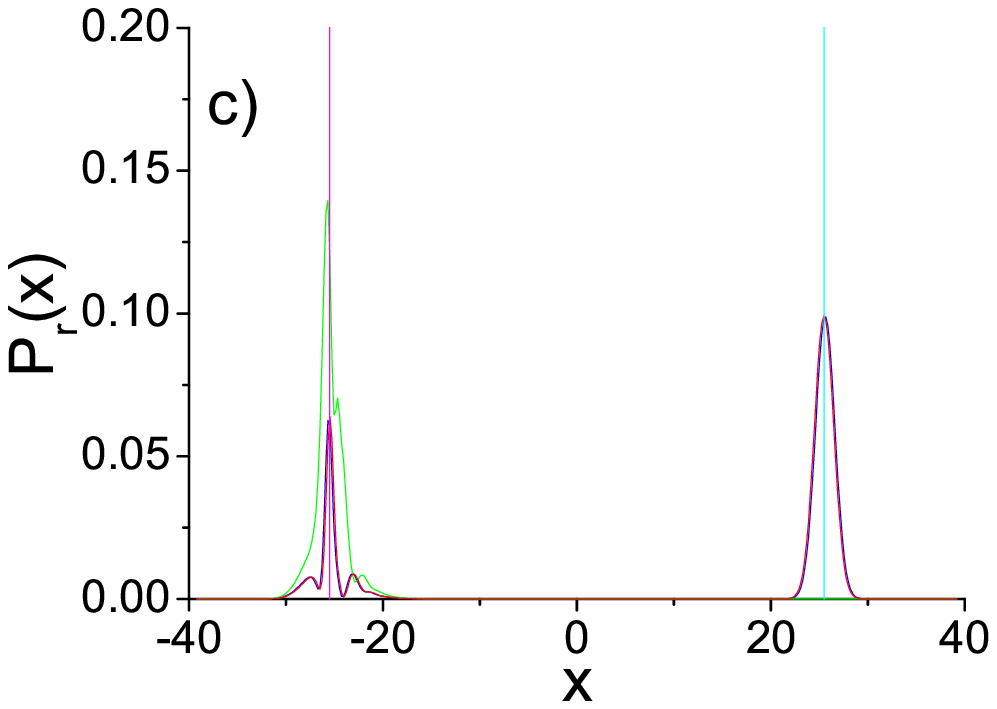,
height=6cm}\epsfig{file=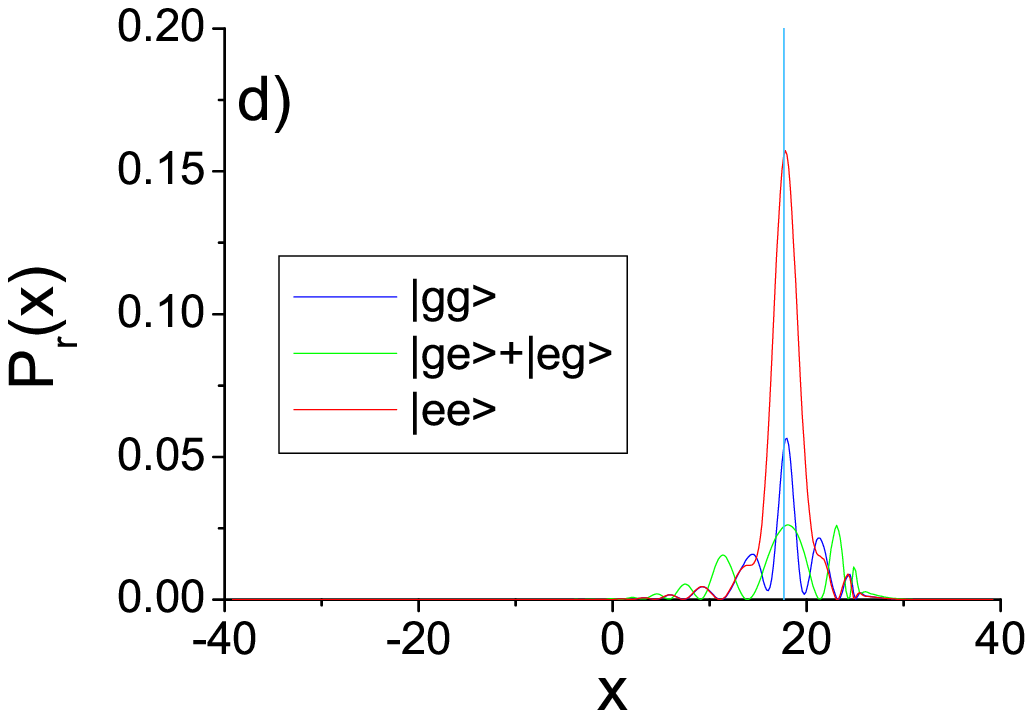, height=6cm} } \caption{The
x-quadrature distribution of the field state after projecting out
the qubits states $r=\ket{gg},\ket{ee}, \ket{ge}+\ket{eg}/\sqrt{2}$
for a series of different times. The vertical lines indicate the
location of the field states as given in equation (\ref{eq:Phik}).
 (a) $t=\pi/(2\omega)$: The x-distribution close to
$t=0$, when the peaks are not well separated, (b)
$t=3\pi/(2\omega)$: After a period of time the sub-distributions,
evolving with different frequencies, separate and can easily be
distinguished, (c) $t=t_{r}/2$: The x-distribution at the first
revival peak, (d) $t=t_{r}$: The x-distribution at the second main
revival peak. The vertical lines indicate $x(t)$ for harmonic
oscillators with frequencies $\omega, \omega\pm
\lambda/\sqrt{\bar{n}}$. For all the diagrams $\lambda=\omega=1$ and
$\bar{n}=200$.\label{fig:xdia}}
\end{figure}

After gaining a clearer understanding of the physics by taking the
large-$\bar{n}$ limit, we now return to our exact calculations based
on equation (\ref{eq:ansoln}). Whilst the Q-function shows a fuller
understanding of the state at a specific time, we shall study an
experimentally more accessible distribution defined as follows.
Evidently the eigenvalues of the operator $\hat{x}=(a+a^\dag)$ are
related to the position variable (electric field) of the harmonic
oscillator which describes the cavity mode. In order to analyze the
behaviour of this system as a function of time, we can plot the
corresponding $x$-distribution of the field state after we have
projected out the qubit states, i.e. we plot
\begin{equation} \label{eq:probdens}
P_r(x,t)=\bra{x,r}\rho(t)\ket{x,r}
\end{equation}
 where $r=gg, ee, eg, ge$ (figures
\ref{fig:xdia}(a) - \ref{fig:xdia}(d)). A convenient numerical
method for obtaining this distribution is to project out a
particular qubit state, transform the density matrix into the
$x$-basis and take the diagonal elements. Obviously, the finite
number of basis states in the calculation will lead to a discrete
set of position eigenvalues $x_i$ for $\hat{x}$, and hence a
discrete set of probabilities $P_r(x_i,t)$, but this set of points
can be made sufficiently dense to form a probability density, as
given in equation (\ref{eq:probdens}).

Although the $Q(\alpha,t)$ function studied by Meunier \emph{et al.}
\cite{Meunier} gives a fuller account of the cavity field than the
above $P_r(x_i,t)$, nevertheless, as shown in figure \ref{fig:xdia},
the latter also captures the salient features of a very interesting
quantum state at hand. In the interest of clarity the pictures in
figure \ref{fig:xdia} are at the same times as the pictures in
figure \ref{fig:qblobs}.

As in figures \ref{fig:qblobs}(a) - \ref{fig:qblobs}(d) we can see
that although the distributions initially overlap, they evolve with
different frequencies, so the peaks begin to separate (figure
\ref{fig:xdia}(b)). After a period of time the distributions overlap
and cause the two revival peaks (figures \ref{fig:xdia}(c) and
\ref{fig:xdia}(d))

\section{Entanglement protocol}

The basis behind the proposed entanglement procedure is a
measurement of the quantum field at some chosen time, which projects
the state of the two qubits into an interesting and useful entangled
state, heralded by the outcome of the field measurement. In quantum
optics the idea of measuring one part of an entangled system to
learn something about the other goes back over twenty years
\cite{imoto85,mil84}. Measurement of a coherent state entangled with
a photon can be used to project the photon state without absorption
of the photon. In our case there is a tripartite entangled system
and projective measurement of the coherent state is used to project the
remaining bipartite qubit system. The relevant measurement is
balanced homodyne detection, where the coherent state is mixed with
a strong local oscillator field on a 50:50 beam-splitter before
photo-detection of the two outgoing fields. It is well known, e.g.
\cite{tyc04}, that in the strong local oscillator limit the
photon counting measurements made on the two outgoing modes
of the beam-splitter
correspond to a projector (e.g. $\ket{x} \bra{x}$) onto a chosen quadrature of
the initial coherent state field being measured, where the
quadrature is set by the chosen phase of the local oscillator field.
As will be seen for our system, at certain times a projection onto
the $x$ quadrature of the field entangled with the qubits can leave
the qubits in an entangled state. For the case of superconducting
qubits coupled to a microwave field mode \cite{Blais,spil06} this requires
homodyne detection at microwave frequencies, rather than the
familiar beam-splitter and photo-detection systems employed at
optical frequencies. The mixing (of signal and local oscillator) and
detection at microwave frequencies, in order to project onto a
defined quadrature of a microwave coherent state field, may be
achievable through use of a single electron transistor (SET), as has
been discussed in detail in \cite{sar05}.

We stress that in our entanglement protocol, the idea is to introduce,
or turn on, the homodyne measurement applied to the cavity field
at some chosen time. This would require some sort of fast (on the
time-scale of the qubit/field evolution discussed in the previous section)
opening or gating of the cavity, to enable the microwave field to be subject
to measurement at a chosen time. Given the long coherence times
(or, correspondingly, high quality factors) observed in
superconducting qubit and microwave mode experiments
\cite{Wallraff,Wallraff2,Schuster}, we neglect decoherence in the
field mode and qubits prior to the measurement, but explicitly
allow for the decoherence of the actual measurement process, through its
projection of the state of the qubits and field.

So, taking homodyne measurement to act as a projective measurement of
$x$, our entanglement procedure is as follows. At a given time $t$
the $x$-quadrature measurement will give a value of $x$ taken from
the distribution $P(x,t)=\sum\limits_r P_r(x,t)$. Repeated
observations would measure the whole distribution $P(x,t)$. Now if
the peaks in $P_r(x,t)$ are well separated for different qubit
states $r$, then the result of a single $x$-measurement indicates
which sub-distribution $P_r(x,t)$ the system is in, and thus the
value of $r$ for that single measurement. Hence the qubit state $r$
is {\it conditioned} on the result of the measurement of $x$. If this
state is an entangled one then we have created an entangled state
out of the unentangled initial condition $\ket{gg}$ by making the
measurement. Furthermore, the creation of the entangled state,
although probabilistic, is heralded on the result of the
$x$-measurement, i.e. the result of the measurement tells us which
peak of the distribution we are in, and hence whether or not we were
successful in our attempt to create the state. Equations
(\ref{eq:productstates})-(\ref{eq:Phik}) show that if the
measurement result corresponds to the field state
$|\Phi_0(t)\rangle$, the qubit state will be a superposition of the
states $\ket{gg}$ and $\ket{ee}$, with a relative phase given by
$\phi=2(\theta+\pi/2+\omega t)$. Although this state is maximally
entangled for any given phase, a two-qubit state with unknown phase
has no extractable entanglement and can be regarded as mixed. Thus
we specify a particular \emph{target state} with a chosen phase and
try to produce this given state.

\begin{figure}\center{
\epsfig{file=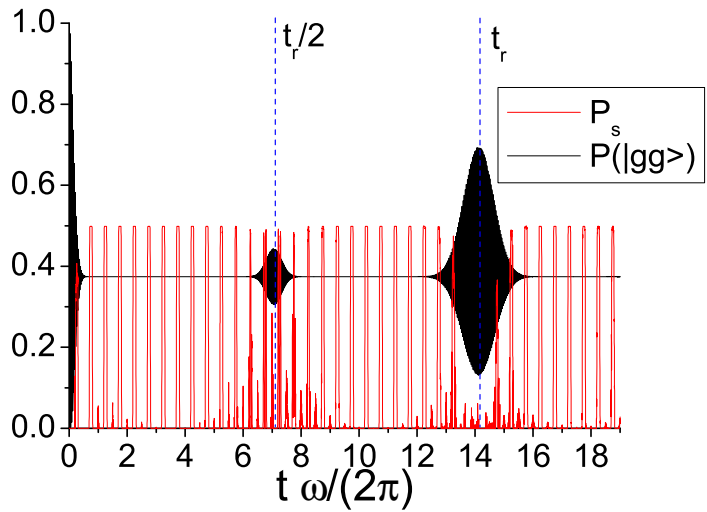, height=8cm}} \caption{Revival of the
qubits' initial state and the probability the entanglement procedure
is successful, $P_s$. The black curve shows the probability that the
qubits are in the state $\ket{gg}$, tracing over the resonator. The
(lighter) red curve shows the probability of obtaining the state
$(\ket{gg}+\ket{ee})/\sqrt{2}$ with a fidelity $F\ge F_{min}$,
$P_s$, as a function of time. The vertical dashed lines indicate the
revival and sub-revival times at $t_r=2\pi\sqrt{\bar{n}}/\lambda$
and $t_r/2$ respectively. We have chosen $\lambda=\omega=1$,
$\bar{n} =200$ and a minimum fidelity $F_{min}=0.9$.
\label{fig:revivalandps}}
\end{figure}

\begin{figure}\centering{
\epsfig{file=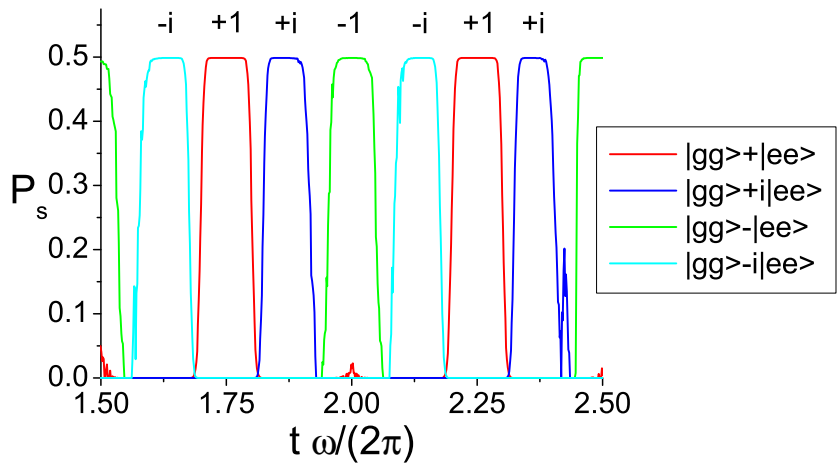, width=12cm} \caption{The probability,
$P_s$, of achieving a fidelity greater than 0.9 for target states of
the form $(\ket{gg}+e^{i\phi}\ket{ee})/\sqrt{2}$ with phases
$0,\pi,\pm\pi/2$. The coupling is given by $\lambda=\omega=1$, and
$\bar{n}=200$. The curves all reach a constant value close to 0.5
for a finite period. The value of $\exp(i\phi)=\pm1,\pm i$ for the
state produced is indicated above each peak.}\label{fig:pspmi}}
\end{figure}

We would like to know how efficient this protocol is at producing a
given target state. We can ask this question in the following way:
if a projective measurement is made on $x$, what is the probability
that the state of the qubits after the measurement will have a
fidelity (given by $F=|\langle a | b\rangle|^2$ for two pure states
$\ket{a}$ and $\ket{b}$) greater than $F_{min}$ with our target
state? We call this value the probability of success, $P_s$.
\begin{eqnarray}
P_s(t)= \sum\limits_i P(x_i,t) \Theta(F-F_{min})
\end{eqnarray}
where $\Theta(F-F_{min})$ is zero for $F\leq F_{min}$ and unity for
$F>F_{min}$. This criterion both allows us to specify \emph{how
good} our state is and tells us \emph{how often} the procedure
works. In figures \ref{fig:revivalandps} and \ref{fig:pspmi} $P_s$
is plotted for the maximally entangled state
$(\ket{ee}+\ket{gg})/\sqrt{2}$ as a function of time, with a minimum
fidelity $F_{min}= 0.9$.

We find that $P_s$ shows a series of flat-topped peaks as a function
of time, figure \ref{fig:revivalandps}, occurring at twice the
resonator frequency $\omega$, indicating that the there is a
significant chance that the measurement will produce the state
$(\ket{ee}+\ket{gg})/\sqrt{2}$ twice every oscillation cycle. Note
that the probability of producing this state is quite large (close
to 0.5) for a significant period of time. We obtain the state
$(\ket{ee}+\ket{gg})/\sqrt{2}$ when the measurement of $x$
corresponds to the peak with frequency $\omega$ (the central peak in
figure \ref{fig:xdia}b). The peaks are most widely separated at the
time $t_r/4$, but it is worth emphasising that we observe peaks in
$P_s$ close to 0.5 as soon as the peaks do not overlap, i.e. at a
much shorter time than the revival.

At other points in the oscillation cycle, the qubits will have a
different relative phase. In figure \ref{fig:pspmi} we have also
plotted the probability of obtaining the states
$(\ket{ee}+e^{-i\phi}\ket{gg})/\sqrt{2}$ with $\phi=0, \pi,
\pm\pi/2$. We can therefore produce any of the maximally entangled
states of the form $(\ket{gg}+e^{-i\phi}\ket{ee})/\sqrt{2}$ by
making the projective measurement at the appropriate time.

We might expect that as the phase is evolving continuously, we would
only get the desired target state at instantaneous points in time.
The fact that we can still get a non-zero $P_s$ for a finite time is
due to two factors: firstly that the states
$(\ket{gg}+e^{-i\phi}\ket{ee})/\sqrt{2}$ for different $\phi$'s are
not orthogonal, and secondly that the coherent state has a finite
size.

\begin{figure}[h!]\centering{
\epsfig{file=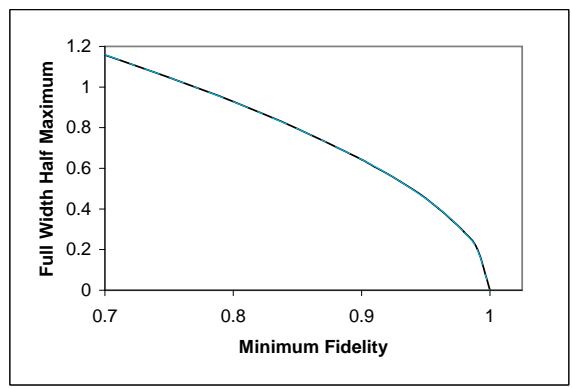, height=5cm} \caption{Full
width at half maximum height for the $P_{s}$ peak, as a function of
fidelity. The numerically calculated value for $\bar{n}=300$ (light
blue dashed line) is compared to the analytical form in equation
(\ref{eq:idealwidth}) (black line). We get good agreement between
these even for values of $\bar{n}$ where the peak shows significant
broadening because the shape of the peak (figure
\ref{fig:fidofnbar}) changes in such a manner that the FWHM value is
relatively unaffected.}\label{fig:halfheight}}
\end{figure}

States with different values of $\phi$ are not in general
orthogonal. Of course, if we were to require that we obtain our
target state with unit fidelity, then we would expect to obtain our
target state only at a single moment in time, and indeed the peaks
in $P_s$ get narrower as $F_{min}$ is increased. However, if we
choose an $F_{min}<1$ which we consider ``good enough,'' then a range
of values of $\phi$ will have fidelity greater than this with the
target state. This leads to a simple trigonometric relationship
between the width of the peak in $P_s$ and the desired fidelity,
\begin{equation} \label{eq:idealwidth}
\Delta t^\infty = \frac{\arccos(2 F_{min}-1)}{\omega},
\end{equation}
which is shown in figure \ref{fig:halfheight}.
In the limit $\bar{n}\to \infty$, we find that $P_s$ as a function
of time is a series of top-hat functions with a width given by
equation (\ref{eq:idealwidth}) and a height of 0.5. The projective
measurement
 ``perfectly" produces a state of the form
$(\ket{gg}+e^{-i\phi}\ket{ee})/\sqrt{2}$ and the width of the peaks
in $P_s$ is solely due to the finite overlap between this state and
the target state.

\begin{figure}[h!]\centering{
\epsfig{file=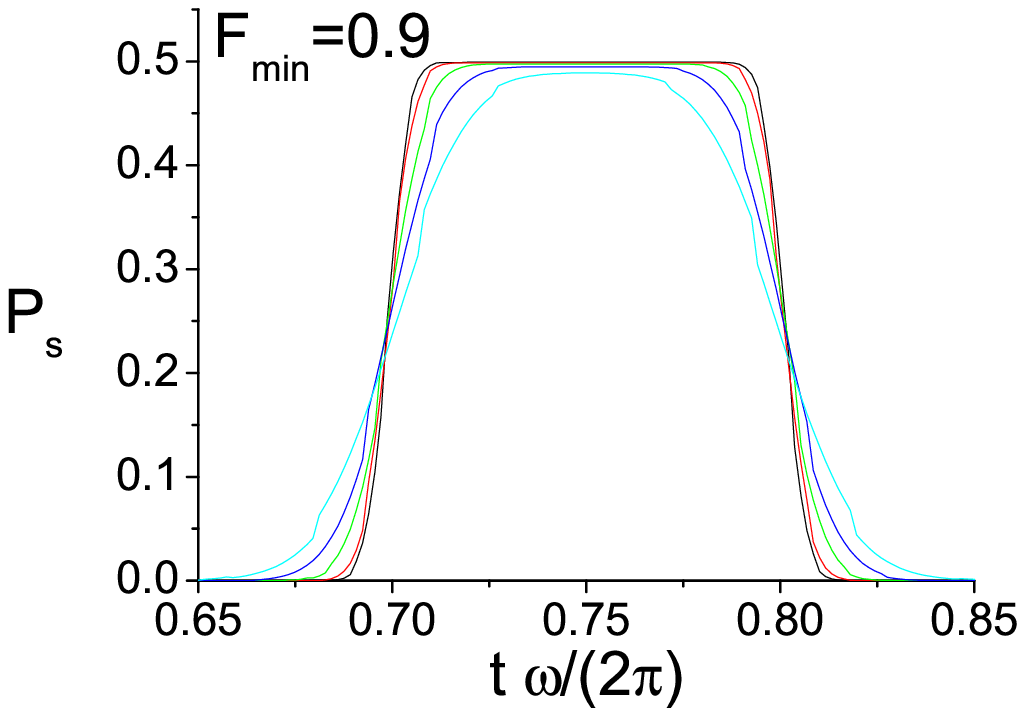, height=4cm}
\epsfig{file=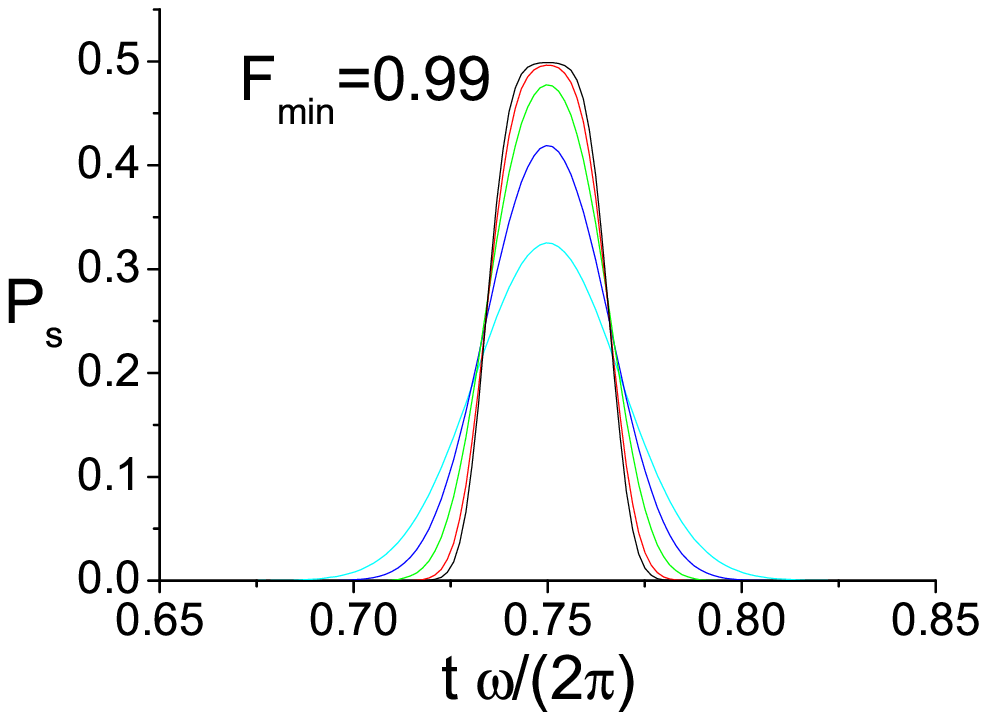, height=4cm}  \caption{The
probability, $P_s$ of achieving desired fidelity $F_{min}$ as a
function of time over one peak, plotted for coherent states with
$\bar{n}=300,200,100,50,25$ (top to bottom). The two sub-plots show
$P_s$ for different values of $F_{min}$. For large coherent states
and low $F_{min}$ the peaks approach a top-hat form. For smaller
coherent states or larger $F_{min}$ the states become more
rounded.}\label{fig:fidofnbar}}
\end{figure}

For finite $\bar{n}$ coherent states we see in figure
\ref{fig:fidofnbar} that the value of $P_s$ changes continuously,
rather than as a top-hat function. This smoothing is due to the fact
that the coherent state has a finite width. If the centre of the
peak in the $x$-distribution corresponds to the state
$(\ket{gg}+e^{-i\phi}\ket{ee})/\sqrt{2}$ the leading and trailing
edges of the peak correspond to qubit states with phases slightly
above or below $\phi$. This means that even when the centre of the
peak has a fidelity lower than $F_{min}$, the leading or trailing
edge may have the target value of $\phi$, leading to peaks in $P_s$
with a greater width, as can be seen in figure \ref{fig:fidofnbar}.
The difference in phase across the distribution also means that when
the centre of the peak has the desired phase, the edges of the peak
have the ``wrong" phase and $P_s$ will be reduced (observable as a
reduction of the step height in figure \ref{fig:fidofnbar}). As
$\bar{n}$ is increased, the difference in phase between the edges
and centre of the peaks disappears, so that for $\bar{n}\to \infty$,
the whole of the peak corresponds to the exact state. Numerical
calculations indicate that the width of the peak is approximately
given by the ``ideal" width $\Delta t^\infty$, plus a term due to
the finite size of the coherent state,
\begin{equation}\label{eqn:quartwid}
\Delta t_{FWQM}=\frac{K(F_{min})}{\sqrt{\bar{n}}}+\Delta t^{\infty}
\end{equation}
where $K(F_{min})$ is a constant that depends on $F_{min}$.

%In figure \ref{fig:maxheight} we plot the height of the peaks as a
%function of $\bar{n}$ for different fidelities.
%
%
%
%\begin{figure}[h!]\centering{
%\epsfig{file=maxheight.eps, height=4cm} \caption{The maximum value
%of the peak for $P_{s}$ as a function of $\bar{n}$ for different
%fidelities. It can be seen that for large values of $\bar{n}$ the
%maximum reached saturates to a value of 0.5. When the fidelity is
%increased, the height of the peak decreases for larger values of
%$\bar{n}$.}\label{fig:maxheight}}
%\end{figure}

There are some points in the evolution that do not lead to a high
value of $P_s$ for any phase $\phi$. This can be observed in figure
\ref{fig:pspmi}, where the peaks in $P_s$ for the states with phases
$\phi=0,\pm \pi/2$ overlap (indicating a continuous change from one
state to the other), but $P_s$ is zero for a finite period between
the peaks representing states with $\phi=\pi$ and $\phi=\pm\pi/2$.
These periods, which occur four times a cycle, represent the times
when two of the $x$-distribution peaks overlap and so the qubit
state produced by the $x$-measurement is not of the form
$(\ket{gg}+e^{-i\phi}\ket{ee})/\sqrt{2}$.

In the limit of $\bar{n} \to \infty$, the peaks in $P_{s}$ are step
functions of height 0.5 with a width determined by equation
(\ref{eq:idealwidth}), which means we can achieve any desired
fidelity with a probability of close to 0.5 as long as we can
measure $x$ at an exact time. Of course, we do not have
instantaneous measurements, and so, depending on the desired
fidelity and the time resolution of the measurement, it may be
desirable to have a smaller coherent state. As $F_{min}$ increases,
we see that the width of the step function decreases (figure
\ref{fig:halfheight}), meaning that the measurement must occur
within an increasingly specific window in time. If this becomes a
limiting factor, it may be preferable to use a smaller coherent
state, increasing the window in which a high fidelity state can be
found at the expense of reducing the probability of obtaining that
state. An additional problem in any real measurement will be
imprecise in space as well as time, i.e. the value of $x$ returned
by the measurement is probabilistic. It is clear that if the
imprecision in the measurement is much smaller than the widths of
the peaks in the $x$-distribution, then the imprecision will have
little effect, and we can treat the measurement as projective.
However, we also find that as long as the imprecision is smaller
than the separation between peaks, the protocol still produces high
fidelity states, although the effect of finite $\bar{n}$ in reducing
the fidelity becomes more pronounced.

It should also be noted that this technique can easily be scaled up
to entangle larger numbers of qubits. For larger numbers of qubits
the entanglement of the states created is not as easy to quantify,
and it may not be possible to couple them to the resonator with
equal strength, and so further research is required in this area.

\section{Conclusions}

We have described a system of two qubits capacitively coupled to a
superconducting microwave resonator, and shown that we can devise a
protocol for heralded probabilistic production of entangled states
of the qubits. This method only requires an initial preparation of
the system in a product state of the qubit ground states, a coherent
state of the field in the resonator, and the ability to perform a
homodyne measurement on the resonator. This protocol can produce
states with a fidelity as high as desired with a probability
approaching 0.5 in the limit of an infinitely large coherent state.
We have shown that there is a trade-off between the fidelity
desired, the probability of obtaining the desired state and the time
window in which the measurement must be performed. Choosing a
smaller coherent state increases the time window of measurement but
reduces the probability of obtaining the desired fidelity.

We have shown that this protocol is in some sense the dual of the
phenomenon of revival in that both cases rely on the fact that we
can regard the whole system as consisting of several field coherent
states for each qubit state, with the different field states
oscillating with different frequencies. Revival occurs when the time
evolution of these states at these frequencies brings them all into
phase with each other. In contrast the entanglement protocol works
best when the states are well separated in phase.

Being based on measurement of a quantum mode, which is coupled
separately to both qubits, our approach to generating entanglement
contrasts with approaches
where the two qubits, although not interacting directly with each other,
are both coupled directly to the same measurement apparatus or detector.
A solid state qubit example of this is given by Ruskov and Korotkov
\cite{rus03}, for two quantum dots coupled to a point contact, or
two Cooper-pair boxes coupled to a SET.
Our entangling protocol also contrasts with approaches
such as that of Schneider and Milburn\cite{schneider}, where the
common resonator mode is both damped and driven for all times.
In such approaches
feedback and control\cite{wang,sar05}, based on homodyne
measurement, can be utilised to enhance the results. It is possible
that similar application of feedback could help for our approach.
However, even without any enhancement over our present results, it
should be noted that such high fidelity, but probabilistic
entanglement---heralded through the measurement outcome---has good
use in Quantum Information Processing. Given the recent progress
with superconducting qubit experiments, initial investigations of
such entangling approaches should soon be possible. In the longer
term, probabilistic but heralded entanglement generation can be
used\cite{barrett,duan} as a basis for efficient quantum computation
using the cluster state approach\cite{raussendorf}.

%%%%%%%%%%%%%%%%%%%%%%%%%%%%%%%%%%%%%%%%%%%%%%%%%%%%%%%%%%%%%%%%%%%%%

\ack   The work of D.A.R. was supported by a UK EPSRC fellowship,
and C.E.A.J. was supported by UK HP/EPSRC case studentship. BLG would
like to thank the hospitality of the CMS of TU Wien.

%%%%%%%%%%%%%%%%%%%%%%%%%%%%%%%%%%%%%%%%%%%%%%%%%%%%%%%%%%%%%%%%%%%%%

\section*{References}
\bibliography{homodyne_meastsedit2}

\end{document}